\documentclass[10pt]{article}
\usepackage{fullpage,citesort,epsfig,graphics,amsbsy}
\usepackage{psfrag}
\usepackage[normal,small]{caption}
\usepackage{newcommand}
\usepackage{newcommand1}

\begin{document}
\setlength{\captionmargin}{20pt}
\begin{titlepage}
\begin{flushright}
UFIFT-HEP-06-\\
\end{flushright}

\vskip 3cm

\begin{center}
\begin{Large}
{\bf One Loop Gluon Gluon Scattering in Light Cone
Gauge\footnote{Supported in part by the Department of Energy under
Grant No. DE-FG02-97ER-41029.}}
\end{Large}

\vskip 2cm {\large J.Qiu} \vskip0.20cm { Department of Physics,
University of Florida, Gainesville FL 32611}

(\today)

\vskip 1.0cm
\end{center}

\begin{abstract}\noindent
We calculate the one loop amplitudes for the two gluon scattering
process in light cone gauge with fermions and scalars circulating
in the loop. This extends the work of \cite{I,II}, in which only
the gluon circulates the loop. By putting all fields in the
adjoint representation with $N_f=2$, $N_s=6$, the scattering
amplitude of gluon by gluon in the special case of $\mathcal{N}=4$
Super Yang-Mills can be obtained. The massive fermion and scalar
with arbitrary representations are also considered.
\end{abstract}
\vfill
\end{titlepage}
\section{Introduction}
In the previous work of \cite{I,II}, we calculated the one loop
gluon scattering amplitude in the light cone world sheet setup.
The main ingredients of light cone world sheet are explained in
\cite{I} and the references therein. This setup shows us how to
organize the calculation in order to cancel the infamous
$1/q^+$(gauge artificial) divergences in the light cone gauge
propagator. In the cases we have calculated, we were able to show
that the artificial divergences actually cancel in physical
processes, hence any prescription(e.g. principle value,
Mandelstam-Leibbrandt prescription \cite{ML}) should work. For the
helicity conserving amplitudes, infrared divergences were
regulated by cutting off $1/q^+$ at $q^+=0$. Not withstanding this
non-covariant regulation, we still got a covariant probability by
including soft Bremsstrahlung and collinear emission(absorption).

This paper extends the work of \cite{I,II}: in addition to the
gauge field itself, scalar and fermion fields are put into the
loop. This paper is organized as follows: in section 2, we review
the derivation of light cone gauge Feynman rules suitable for
massless fields and the light cone world sheet. In section 3, the
calculational techniques of \cite{II} are briefly reviewed. In
section 4, the results are listed without the calculational
details, since most of calculation is done by computers. When we
specialise to the case $N_g=1$, $N_f=2$ (Dirac fermions) and
$N_s=6$, we find that the gluon scattering amplitude in
$\mathcal{N}=4$ SYM takes on a very simple form.

In the last part of the paper(section 5 and 6), we deal with
massive matter fields. Some of the techniques that are designed
for massless fields in the first part of the calculation are not
useful anymore. So we simply follow the textbook procedure, but
spinors in light cone are extensively used to organize the
calculation(section 5). Section 6 lists the results.

\section{The Light Cone Setup}
The light cone gauge Feynman rules are usually obtained by the
lagrangian method, namely, we first set $A_-=0$ and integrate out
$A^-$. Then the Feynman rules can be read off from the lagrangian,
which is a function of $A^1$ and $A^2$.

Here, a more flexible method is used. The gluon propagator in
light cone gauge is: \bea\frac{-i(g^{\mu\nu}-\frac{g^{\mu
+}k^{\nu}}{k^+}-\frac{g^{\nu
+}k^{\mu}}{k^+})}{k^2+i\epsilon}\label{glue_prop}\eea Note our
metric is $diag(1,-1,-1,-1)$. The numerator can be factored
into:\bea (g^{\mu\nu}-\frac{g^{\mu +}k^{\nu}}{k^+}-\frac{g^{\nu
+}k^{\mu}}{k^+})=-(\epsilon_\vee^{\mu}\epsilon_\wedge^{\nu}+\epsilon_\wedge^{\mu}\epsilon_\vee^{\nu})-\frac{g^{\mu
+}g^{\nu+}}{k^{+2}}k^2\label{glue_prop_decomp}\eea Where
$\epsilon_{\wedge,\vee}$ are light cone gauge polarization vectors
given by \bea \epsilon_{\vee}^{\mu}=\frac{1}{\sqrt2}(\frac{p^1-
ip^2}{p^0+p^3},1,-i,-\frac{p^1-
ip^2}{p^0+p^3})=\frac{1}{\sqrt2}(\frac{p^{\wedge}}{p^+},1,-i,-\frac{p^{\wedge}}{p^+})\textrm{
; }\epsilon_{\wedge}=\epsilon_{\vee}^*
\eea They satisfy $\epsilon\cdot\epsilon^{*}=-1$ and
$k\cdot\epsilon=0$. These polarisation vectors are defined both
on-shell and off-shell.

The Feynman rules are obtained by dotting the polarisation vectors
into the covariant three or four point vertices.
\begin{figure}[!h]
\begin{center}
\psfrag{nu,b,P2}{$\nu,\;b,\;p_2$}\psfrag{mu,a,P1}{$\mu,\;a,\;p_1$}
\psfrag{rho,c,P3}{$\rho,\;c,\;p_3$}
\includegraphics[width=1.2in]{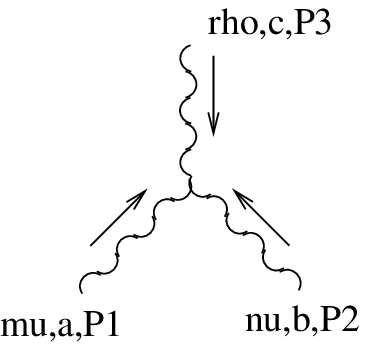}\caption{}
\label{ggg}
\end{center}
\end{figure}
The tri-gluon vertex Fig.\ref{ggg}, for example, becomes  \bea
V_{ggg}=
gf^{abc}(-\epsilon^*_{1\mu})(-\epsilon^*_{2\nu})(-\epsilon^*_{3\rho})[g^{\mu\nu}(p_1-p_2)^\rho+g^{\nu\rho}(p_2-p_3)^\mu+g^{\rho\mu}(p_3-p_1)^\nu]\nn\eea
Setting $\epsilon_1,\ \epsilon_2=\epsilon_{\wedge}$,
$\epsilon_3=\epsilon_\vee$, the above becomes: \bea
gf^{abc}[(p_2-p_3)^+\frac{p_1^\wedge}{p_1^+}-(p_2-p_3)^\wedge+(p_3-p_1)^+\frac{p_2^\wedge}{p_2^+}-(p_3-p_1)^\wedge]=2gf^{abc}\frac{(p_1+p_2)^+}{p_1^+p_2^+}K_{21}^\wedge\eea
where $K_{i,j}^{\mu}:=(p_i^+p_j^\mu-p_j^+p_i^\mu)$. They are
related to spinor products according to:\bea
K_{i,j}^{\vee}=p_i^+p_j^+[p_i|p_j]=p_i^+p_j^+p_i^{\dot{a}}p_{j\dot{a}}\nn\eea
The spinor notation here is also different from the conventional
one \cite{parketaylor}. The reader can refer to the appendix for
an explanation of the spinor notation.

The gluon propagator Eq.\ref{glue_prop} almost factorises into the
product of two polarisation vectors. While the third term on the
rhs of Eq.\ref{glue_prop_decomp} will make an extra contribution
to the four point vertex. For example, consider the t-channel
diagram Fig.\ref{t_chan}, the first two terms of
Eq.\ref{glue_prop_decomp} can be associated with the two tri-gluon
vertices, the third term, which describes the mediation of $A^-$,
gives:
\begin{figure}[!h]
\begin{center}
\psfrag {mu,p1}{$\mu,\;p_1$}\psfrag{nu,p2}{$\nu,\;p_2$}
\psfrag{rho,p3}{$\rho,\;p_3$}\psfrag{sigma,p4}{$\sigma,\;p_4$}
\includegraphics[width=1.5in]{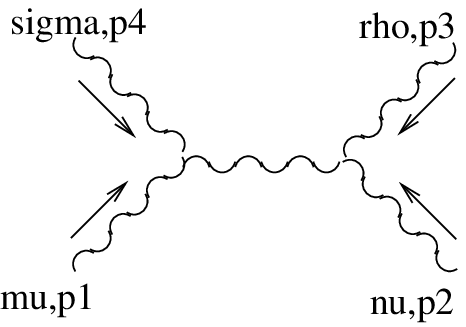}\caption{}
\label{t_chan}
\end{center}
\end{figure}
\bea \epsilon_{1\mu}^*\epsilon_{4\sigma}^*V^{\sigma \mu
\alpha}\frac{i}{(p_1+p_4)^2}\epsilon_{2\nu}^*\epsilon_{3\rho}^*V^{\nu
\rho
\beta}\frac{\delta_\alpha^{+}\delta_\beta^{+}(p_1+p_4)^2}{(p_1^{+}+p_4^{+})^2}=\frac{i\epsilon_1^*\cdot\epsilon_4^*(p_4^+-p_1^+)\epsilon_2^*\cdot\epsilon_3^*(p_2^+-p_3^+)}{(p_1^{+}+p_4^{+})^2}\eea
What's happening here is that the explicit factor of $k^2$ in the
third term of Eq.\ref{glue_prop_decomp} cancels the propagator,
effectively making a four point contact vertex.

\begin{figure}[!h]
\begin{center}
\psfrag {'mu a'}{$\mu,\;a$} \psfrag {'q'}{$q$}\psfrag{'c',
'P3'}{$c,\;p_1$} \psfrag{'b', 'P2'}{$b,\;p_2$}
\includegraphics[width=1.5in]{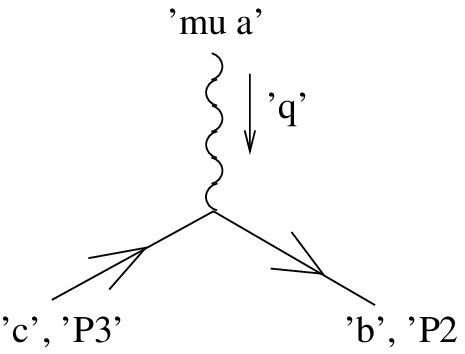}
\label{gff1}
\end{center}
\end{figure}

The fermion-gluon vertex is the usual $ig\,\gamma^\mu\,t^a$. We
can set $\mu=\wedge\;\mathrm{or}\;\vee$ by dotting
$-\epsilon_{\wedge,\vee}^{*},$ into $\mu$.\footnote{ Calling
polarisations by $\vee$ or $\wedge$ is potentially confusing,
especially if you are looking at the diagram up-side-down.
So,sometimes it is clearer to associate $\wedge$ with 'in' and
$\vee$ with 'out'.} Multiplying the spinors to the gamma matrix,
we get (assuming now the fermion is left-handed): \bea &&ig(t^a)_{bc}\sqrt{2p_1^+ p_2^+}(-\epsilon_\wedge^{*\mu})\left[\begin{array}{cc}0&p_2^{\dot{\alpha}}\end{array}\right]\left|\begin{array}{cc}0&(\sigma_\mu)^{\alpha\dot{\alpha}}\\ (\bar{\sigma}_\mu)_{\dot{\alpha}\alpha}&0\end{array}\right|\left[\begin{array}{c}p_1^{\alpha}\\0\end{array}\right]=-ig(t^a)_{bc}\sqrt{2p_1^+p_2^+}(-\sqrt{2})[p_2\,|\eta]\langle q|p_1\rangle\nn\\
&&=-2ig (t^a)_{bc}
\sqrt{p_1^{+}p_2^{+}}\left(\frac{q^\wedge}{q^+}-\frac{p_1^\wedge}{p_1^+}\right)=-2ig
t^a
\frac{\sqrt{p_1^{+}p_2^{+}}}{q^+p_1^+}K^{\wedge}_{p_1,q}\;\rightarrow\;-2ig
t^a \frac{p_2^{+}}{q^+p_1^+}K^{\wedge}_{p_1,q}\label{gff}\eea In
Eq.\ref{gff}, in order to avoid defining what is $\sqrt{p^+}$, we
choose to associate $p_2^+$ instead of $\sqrt{p_1^{+}p_2^{+}}$ to
a vertex. This won't cause any problem, since fermion loop always
closes, so the square root always appears in pairs.

The fermion propagator is given by: $ip_\mu
\gamma^\mu/(p^2+i\epsilon)$. We can decompose $p\cdot\gamma$
according to \bea
&&p\cdot\sigma={\sqrt{2}}\left|\begin{array}{cc}p^-&-p^\wedge\\-p^\vee&p^+\end{array}\right|={\sqrt{2}p^+}\left[\begin{array}{c}-\frac{p^\wedge}{p^+}\\1\end{array}\right]\left|\begin{array}{cc}-\frac{p^\vee}{p^+}&1\end{array}\right|+\left|\begin{array}{cc}\frac{p^2}{\sqrt{2}p^+}&0\\0&0\end{array}\right|=\sqrt{2}p^+|p\,\rangle[p\,|+\frac{p^2}{\sqrt{2}p^+}|\eta\,\rangle[\eta\,|\nn\\
&&p\cdot\bar{\sigma}={\sqrt{2}}\left|\begin{array}{cc}p^+&p^\wedge\\p^\vee&p^-\end{array}\right|={\sqrt{2}p^+}\left[\begin{array}{c}1\\\frac{p^\vee}{p^+}\end{array}\right]\left[\begin{array}{cc}1&\frac{p^\wedge}{p^+}\end{array}\right|+\left|\begin{array}{cc}0&0\\0&\frac{p^2}{\sqrt{2}p^+}\end{array}\right|=\sqrt{2}p^+|p\,]\langle
p\,|+\frac{p^2}{\sqrt{2}p^+}|\eta\,]\langle
\eta\,|\label{fermi_prop}\eea

So the fermion propagator almost factorises too. It is also
possible to contract a pair of vertices here: the second term of
Eq.\ref{fermi_prop} will again cancel a propagator. For example:
\begin{figure}[!h]
\begin{center}
\psfrag {b,p1}{$b,\;p_1$}\psfrag{c,p2}{$c,\;p_2$}
\psfrag{rho,d,p3}{$\rho,\;\wedge,\;p_3$}\psfrag{sigma,a,p4}{$\sigma,\;\vee,\;p_4$}
\includegraphics[width=1.5in]{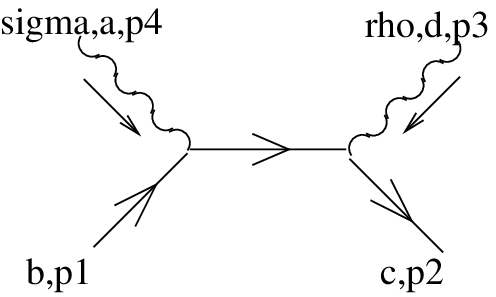}
\end{center}
\end{figure}
\bea&&-g^2(t^d)_{ce}(t^a)_{eb}
\sqrt{2p_1^+p_2^+}\,[p_2|(-\epsilon_3^*\cdot\bar{\sigma})\frac{i(p_1+p_4)\cdot\sigma}{(p_1+p_4)^2}(-\epsilon_4^*\cdot\bar{\sigma})|p_1\rangle\,\sqrt{\sqrt{2}p_1^+}\nn\\
&&\rightarrow-g^2(t^d)_{ce}(t^a)_{eb}\sqrt{2p_1^+p_2^+}[p_2\,|(\sqrt{2}|\eta\,]\langle
p_3\,|)\cdot\frac{i}{(p_1+p_4)^2}\cdot(\frac{(p_1+p_4)^2}{\sqrt{2}(p_1^++p_4^+)}|\eta\,\rangle[\eta\,|)\cdot
(\sqrt{2}|p_4\,]\langle\eta\,|)\,|p_1\,\rangle\nn\\
&&=-2ig^2 (t^d)_{ce}
(t^a)_{eb}\frac{\sqrt{p_1^+p_2^+}}{p_1^++p_4^+}\;\rightarrow\;-2ig^2
(t^d)_{ce} (t^a)_{eb}\frac{p_2^+}{p_1^++p_4^+}\eea

The scalar Feynman rules have no suspense in them at all, and can
be read off from any field theory book. The Feynman rules that
pertain to our calculation will be summarised in the appendix. The
main property of the Feynman rules above is the absence of $p^-$.

\section{Brief Description of Calculational Procedure(massless case)}
\begin{figure}[!h]
\begin{center}
\psfrag {q}{$q$}\psfrag {p1}{$k_0$}\psfrag {p2}{$k_1$}\psfrag
{p3}{$k_2$} \psfrag {1}{$1$}\psfrag {2}{$2$}\psfrag {3}{$3$}
\includegraphics[width=1.1in]{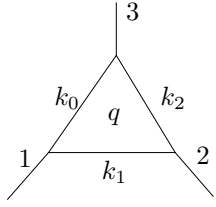}\caption{Here we assume only $(k_0-k_2)^2\ne0$, and $k_2^+>k_1^+>k_0^+$}
\label{sample_cal}
\end{center}
\end{figure}
In the above figure \ref{sample_cal}, the $k$'s and $q$ are the
dual momenta. They are related to the real momenta coming into the
three legs according to: $p_1=k_1-k_0$, $p_2=k_2-k_1$ and
$p_3=k_0-k_2$. The unregulated integrands have a symmetry under
$k_i\to k_i+a$, which would ensure that each diagram depends only
on the real momenta. But here, as in \cite{I,II}, we use a
regulator $\exp{\{-\delta[(q^1)^2+(q^2)^2]\}}$ that breaks this
symmetry. Hence, a regulated amplitude can depend on the
individual dual momenta. This seemingly unwieldy regulator is
designed for the world sheet purpose, which doesn't concern us
here. The result doesn't differ too much when a cut off regulator
is used.

The calculation roughly goes as follows,

1. Exponentiate all propagators according to
\bea\frac{i}{p^2+i\epsilon}=\int_0^{\infty} {dT\,e^{iTp^2}}\nn\eea

For the schematic diagram Fig.\ref{sample_cal}. We have \bea
\Gamma&=&\int \frac{d^4q}{(2\pi)^4}\,\frac{i}{(q-k_0)^2}\frac{i}{(q-k_1)^2}\frac{i}{(q-k_2)^2}\nn\\
&=&\int{\frac{d^4q}{(2\pi)^4}\,d\Uu d\T d\V
e^{i\Uu(q-k_0)^2+i\T(q-k_1)^2+i\V(q-k_2)^2}}\nn\eea

2. Integrate out $q^-$, leaving a delta function relating $q^+$ to
the Feynman parameters(this step requires the absence of $q^-$ in
all Feynman rules). \bea
\Gamma&=&\int{\frac{dq^+d^2\mathbf{q}}{(2\pi)^3}\frac{1}{2}d\Uu
d\T d\V\delta(\sum{T_i
k_i^+}-\sum{T_i}q^+)e^{i\Uu(q-k_0)^2+i\T(q-k_1)^2+i\V(q-k_2)^2}}\nn\eea

3. Integrate $q^1,\, q^2$ using $\exp{[-\delta \mathbf{q^2}]}$ as
a damping factor. \bea \Gamma&=&\int{d\Uu d\T
d\V\frac{dq^+}{(2\pi)^3}\frac{\pi}{2i(\sum{T_i}-i\delta)}\delta(\sum{T_i
k_i^+}-\sum{T_i}q^+)e^{i\Uu\V(k_0-k_2)^2}}\nn\\
&=&\int{\frac{dq^+}{16\pi^2}d\Uu d\T
d\V\delta(\sum{T_i}-1)\delta(\sum{T_i
k_i^+}-q^+)\frac{1}{\Uu\V(k_0-k_2)^2}}\nn\eea

Here is the rub: we cannot simply integrate over $\Uu$,$\T$ and
$\V$, because the prefactor of this diagram will have up to second
order poles at $q^+=k_i^+,\ i=0,1,2$ . In order to show the
cancellation of gauge artificial divergences, we proceed as
follows: first eliminate one Feynman parameter in favour of $q^+$:
\bea
\Pt<q^+<\Po:\;\T=\frac{x(q^+-\Pt)}{\Po-\Pt},\;\;\V=\frac{(1-x)(q^+-\Pt)}{\Pf-\Pt},\;\;\Uu=1-\T-\V\nn\\
\Po<q^+<\Pf:\;\T=\frac{(1-x)(q^+-\Pf)}{\Po-\Pf},\;\;\Uu=\frac{x(q^+-\Pf)}{\Pt-\Pf},\;\;\V=1-\T-\Uu\nn
\eea Now $d\Uu d\T d\V\delta(\sum{T_i}-1)\delta(\sum{T_i
k_i^+}-q^+)=dq^+dxJ$. After integrating out $x$, we are left with
a function of $q^+$ which is defined differently in different
regions: $\Pt<q^+<\Po$ and $\Po<q^+<\Pf$. All these can be
visualised very clearly when we represent a Feynman diagram on the
light cone world sheet. The details can be found in \cite{I,II}.

Our observation is that: in each region, all poles
cancel.\footnote{in the case helicity conserving amplitude, all
poles cancel up to infrared terms, but since infrared divergence
is always proportional to a tree, they are easy to recognize and
deal with.} Hence we can perform the final $q^+$ integral and
obtain the results.
\section{Results}
Since the Feynman rules of fermions and scalars are similar to the
pure gluon case, the evaluation of scalar and fermion loop is
similar to \cite{II}. In fact, for the helicity conserving
amplitudes, the reader needs only work out the prefactor of each
diagram (as a rational function of $q^+$) and read off the results
from \cite{II}. All of the calculation is done by MATLAB program,
so we only list the results here.

We will be following \cite{Mangano} and decompose an n-particle
amplitude into: \bea
\mathcal{M}_n=\sum_{perm'}{\Tr(t^{a_1}t^{a_2}...t^{a_n})M(p_1,\epsilon_1;p_2,\epsilon_2;...;p_n,\epsilon_n)}\eea
where $perm'$ is over non cyclic permutations for complex
representation, non cyclic and non reflexive permutations for real
representations.
\subsection{Self-energy Diagrams}
\begin{figure}[!h]
\begin{center}
\psfrag {'mu','a'}{$\wedge,\;a$}\psfrag {'nu','b'}{$\wedge,\;b$}
\psfrag {'q'}{$q$}\psfrag {p}{$p$}\psfrag{'k1'}{$k_1$}
\psfrag{'k0'}{$k_0$}
\includegraphics[width=1.5in]{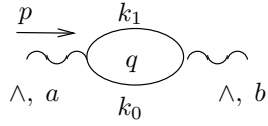}\caption{A helicity violating self-mass diagram}
\label{sf_mass}
\end{center}
\end{figure}
The helicity violating self-energy diagram is given by: \bea
&&\Pi^{\wedge\wedge}_S=\frac{1}{3}\frac{ig^2}{16\pi^2}\Tr[t^at^b]\left[k_0^{\wedge2}+k_0^\wedge k_1^{\wedge}+k_1^{\wedge2}\right]\nn\\
&&\Pi^{\wedge\wedge}_F=\frac{-4}{3}\frac{ig^2}{16\pi^2}\Tr[t^at^b]\left[k_0^{\wedge2}+k_0^\wedge k_1^{\wedge}+k_1^{\wedge2}\right]\nn\\
&&\Pi^{\wedge\wedge}_G=\frac{2}{3}\frac{ig^2}{16\pi^2}\Tr[t^at^b]\left[k_0^{\wedge2}+k_0^\wedge
k_1^{\wedge}+k_1^{\wedge2}\right]\eea This diagram describes the
amplitude of a left handed gluon flip into a right handed gluon.
It is only nonzero because the regulator we had used doesn't
respect Lorentz invariance. This is purely an artifact, and it has
to be cancelled by a counter term. Also, we observe that
$\Pi^{\wedge\wedge}_S\times N_s+\Pi^{\wedge\wedge}_F\times
N_f+\Pi^{\wedge\wedge}_G\times N_g=0$ in the $\mathcal{N}=4$ SYM
case.

\begin{figure}[!h]
\begin{center}
\psfrag {'mu','a'}{$\wedge,\;a$}\psfrag {'nu','b'}{$\vee,\;b$}
\psfrag {'q'}{$q$}\psfrag {p}{$p$}\psfrag{'k1'}{$k_1$}
\psfrag{'k0'}{$k_0$}
\includegraphics[width=1.5in]{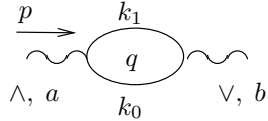}\caption{A helicity conserving self-mass diagram}
\end{center}
\end{figure}
\bea
&&\Pi^{\wedge\vee}_S=-\frac{ig^2}{16\pi^2}\Tr[t^at^b]\left[-\frac{1}{\delta}+\int_0^1{dx
x(1-x)p^2\log{x(1-x)p^2\delta e^\gamma}}\right]\nn\\
&&\Pi^{\wedge\vee}_F=-2\frac{ig^2}{16\pi^2}\Tr[t^at^b]\left[\frac{2}{\delta}+\int_0^1{dx\left[x^2+(1-x)^2\right]p^2\log{x(1-x)p^2\delta
e^\gamma}}\right]\nn\\
&&\Pi^{\wedge\vee}_G=-2\frac{ig^2}{16\pi^2}\Tr[t^at^b]\left[-\frac{1}{\delta}+\int_0^1{dx\left[\frac{x}{1-x}+\frac{1-x}{x}+x(1-x)\right]p^2\log{x(1-x)p^2\delta
e^\gamma}}\right]\label{sfmass_massless}\eea In the above list, we
have omitted part of the gluon energy terms that is non-covariant
(proportional to $p^+$), which was interpreted as a world sheet
boundary cosmological constant in \cite{I}. \footnote{In the light
cone setup, certain tadpole diagrams are ill defined, and are
discarded from the beginning. This is the reason of the
non-covariance in the result.}

\begin{figure}[!h]
\begin{center}
\psfrag {a}{a}\psfrag {b}{b}\psfrag {c}{c}\psfrag {d}{d} \psfrag
{e}{e}\psfrag {f}{f}
\includegraphics[width=1.5in]{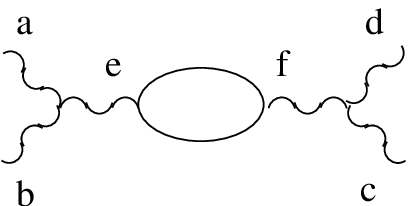}\caption{}
\label{sf_factor2_explain}
\end{center}
\end{figure}

When we try to fit the self-energy diagrams into the big picture
as in Fig.\ref{sf_factor2_explain},
the trace factor becomes (assuming that $t^a$ is in a real
representation): \bea f^{abe}\Tr[t^e
t^f]f^{fcd}&=&\Tr[(-i)[t^a,t^b](-i)[t^c,t^d]]=-\Tr[t^at^bt^ct^d]+\Tr[t^at^bt^dt^c]+\Tr[t^bt^at^ct^d]-\Tr[t^bt^at^dt^c]\nn\\
&&\rightarrow -2\Tr[t^at^bt^ct^d]\eea

\subsection{Triangle Diagrams}
\begin{figure}[!h]
\begin{center}
\psfrag {'mu','a','p1'}{$\wedge,\;a,\;p_1$}\psfrag
{'nu','b','p2'}{$\wedge,\;b,\;p_2$}\psfrag
{'rho','c','p3'}{$\vee,\;c,\;p_3$}\psfrag {'k1'}{$k_1$}\psfrag
{'k2'}{$k_2$}\psfrag {'k0'}{$k_0$}\psfrag {'q'}{$q$}
\includegraphics[width=1.7in]{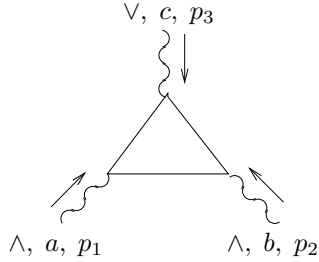}\caption{Triangle diagram}
\label{trig}
\end{center}
\end{figure}
The result for the triangle diagram Fig.\ref{trig} is:\bea
\Gamma^{\wedge\wedge\vee}_S&=&\frac{ig^3}{8\pi^2}\Tr[t^at^bt^c]\frac{-2p_3^+}{p_1^+p_2^+}K^\wedge_{2,1}\left[\frac{1}{6}\log{p_o^2\delta
e^{\gamma}}-\frac{1}{9}-\alpha\frac{1}{6}\frac{p_1^+p_2^+}{p_3^{+2}}\right]\nn\\
\Gamma^{\wedge\wedge\vee}_F&=&-\frac{ig^3}{4\pi^2}\Tr[t^at^bt^c]\frac{-2p_3^+}{p_1^+p_2^+}K^\wedge_{2,1}\left[-\frac{2}{3}\log{p_o^2\delta
e^{\gamma}}+\frac{16}{9}-\alpha\frac{1}{3}\frac{p_1^+p_2^+}{p_3^{+2}}\right]\nn\\
\Gamma^{\wedge\wedge\vee}_G&=&\frac{ig^3}{8\pi^2}\Tr[t^at^bt^c]\frac{-2p_3^+}{p_1^+p_2^+}K^\wedge_{2,1}\left[-\frac{11}{3}\log{p_o^2\delta
e^{\gamma}}+\frac{70}{9}-\alpha\frac{1}{3}\frac{p_1^+p_2^+}{p_3^{+2}}\right]+\textrm{infrared
terms}\eea where $\alpha=1$ if leg 3 is off shell and 0 otherwise,
and $p_o$ is the off shell momentum. The so called infrared terms
present in the gluon triangle diagram will eventually be combined
with infrared terms from other diagrams to become proportional to
a tree. They are given in \cite{II} and are not repeated here.

All of the above also contain an anomalous term:

\bea
&&\Gamma_S^{ano}=\frac{ig^3}{8\pi^2}\Tr[t^at^bt^c]\frac{1}{3}(k_0^\wedge+k_1^\wedge+k_2^\wedge)\nn\\
&&\Gamma_F^{ano}=-\frac{ig^3}{4\pi^2}\Tr[t^at^bt^c]\frac{2}{3}(k_0^\wedge+k_1^\wedge+k_2^\wedge)\nn\\
&&\Gamma_G^{ano}=\frac{ig^3}{8\pi^2}\Tr[t^at^bt^c]\frac{2}{3}(k_0^\wedge+k_1^\wedge+k_2^\wedge)\nn\eea
here the $k$'s are dual momenta, they arise because $\delta$ is an
exponential damping factor of the transverse dual momenta. Had we
used a cut off regulator ($(\mathbf{q}-\mathbf{k}_1)^2<\Lambda$
for example) instead of $\delta$, the anomalous terms would change
according to:$
(\mathbf{k}_0+\mathbf{k}_1+\mathbf{k}_2)\;\rightarrow\;(\mathbf{k}_0-\mathbf{k}_1)+(\mathbf{k}_2-\mathbf{k}_1)$.
These polynomial terms must be cancelled by counter terms.

The MHV triangles (with all three legs having the same helicity) give: \bea\Gamma^{\wedge\wedge\wedge}_S&=&\frac{ig^3}{8\pi^2}\Tr[t^at^bt^c]\frac{(K^{\wedge}_{21})^3}{p_1^+p_2^+p_3^+}\bigg[\frac{2}{3p_o^2}\bigg]\nn\\
\Gamma^{\wedge\wedge\wedge}_F&=&-\frac{ig^3}{4\pi^2}\Tr[t^at^bt^c]\frac{(K^{\wedge}_{21})^3}{p_1^+p_2^+p_3^+}\bigg[\frac{4}{3p_o^2}\bigg]\nn\\
\Gamma^{\wedge\wedge\wedge}_G&=&\frac{ig^3}{8\pi^2}\Tr[t^at^bt^c]\frac{(K^{\wedge}_{21})^3}{p_1^+p_2^+p_3^+}\bigg[\frac{4}{3p_o^2}\bigg]\eea

\subsection{Scattering Amplitude}
Finally, we simply list the results for the gluon scattering
amplitude with massless fermions and scalars.

The tree level amplitude for four gluons with the same helicity is
zero. At one loop, the amplitude is
\bea &&A^{\wedge\wedge\wedge\wedge}_S=\frac{ig^4}{6\pi^2}\Tr[t^at^bt^ct^d]\frac{K_{43}^{\wedge}K_{32}^{\wedge}K_{21}^{\wedge}K_{14}^{\wedge}}{p_1^+p_2^+p_3^+p_4^+st}\nn\\
&&A^{\wedge\wedge\wedge\wedge}_F=\frac{-2ig^4}{3\pi^2}\Tr[t^at^bt^ct^d]\frac{K_{43}^{\wedge}K_{32}^{\wedge}K_{21}^{\wedge}K_{14}^{\wedge}}{p_1^+p_2^+p_3^+p_4^+st}\nn\\
&&A^{\wedge\wedge\wedge\wedge}_G=\frac{ig^4}{3\pi^2}\Tr[t^at^bt^ct^d]\frac{K_{43}^{\wedge}K_{32}^{\wedge}K_{21}^{\wedge}K_{14}^{\wedge}}{p_1^+p_2^+p_3^+p_4^+st}\eea
The tree level amplitude with only one unlike helicity is zero
too. At one loop level, the amplitude is \bea
&&\hspace{.3in}A^{\wedge\wedge\wedge\vee}_S=
\frac{ig^4}{48\pi^2}\Tr[t^at^bt^ct^d](s+t)\frac{K_{13}^{\wedge
2}p_2^+p_4^+}{K_{43}^{\wedge}K_{32}^{\vee}K_{21}^{\vee}K_{14}^{\wedge}}\nn\\
&&\hspace{.3in}A^{\wedge\wedge\wedge\vee}_F=
\frac{-ig^4}{12\pi^2}\Tr[t^at^bt^ct^d](s+t)\frac{K_{13}^{\wedge
2}p_2^+p_4^+}{K_{43}^{\wedge}K_{32}^{\vee}K_{21}^{\vee}K_{14}^{\wedge}}\nn\\
&&\hspace{.3in}A^{\wedge\wedge\wedge\vee}_G=\frac{ig^4}{24\pi^2}\Tr[t^at^bt^ct^d](s+t)\frac{K_{13}^{\wedge
2}p_2^+p_4^+}{K_{43}^{\wedge}K_{32}^{\vee}K_{21}^{\vee}K_{14}^{\wedge}}
\eea The helicity conserving amplitude is non-zero at tree level.
They are given by \cite{parketaylor}: \bea
&&A^{\wedge\wedge\vee\vee}_{tree}=ig^2f^{abe}f^{ecd}\frac{-2K_{12}^{\wedge
4}p_3^+p_4^+}{K_{43}^{\wedge}K_{32}^{\wedge}K_{21}^{\wedge}K_{14}^{\wedge}p_1^+p_2^+}\nn\\
&&A^{\wedge\vee\wedge\vee}_{tree}=ig^2f^{abe}f^{ecd}\frac{-2K_{13}^{\wedge
4}p_2^+p_4^+}{K_{43}^{\wedge}K_{32}^{\wedge}K_{21}^{\wedge}K_{14}^{\wedge}p_1^+p_3^+}\eea
The factor $f^{abe}f^{ecd}$ can be converted to
$-1/C(G)\Tr\left[[t^a,t^b][t^c,t^d]\right]\,\to\,-2/C(G)\Tr\left[t^at^bt^ct^d\right]$.

At one loop level, the amputated Green's function is (with
infrared terms omitted) \noindent\bea
A^{\wedge\wedge\vee\vee}_S&=&\frac{ig^4}{8\pi^2}\Tr[t^at^bt^ct^d]\bigg\{\frac{-2K_{12}^{\wedge
4}p_3^+p_4^+}{K_{43}^{\wedge}K_{32}^{\wedge}K_{21}^{\wedge}K_{14}^{\wedge}p_1^+p_2^+}\bigg[\frac{1}{18}+\frac{1}{6}\log{\delta
e^\gamma{t}}
\bigg]-\frac{1}{6}{\textrm{\Large{$\times$}}}+\frac{1}{3}\bigg\}\nn\\
\nn\\
A^{\wedge\wedge\vee\vee}_F&=&\frac{-ig^4}{4\pi^2}\Tr[t^at^bt^ct^d]\bigg\{\frac{-2K_{12}^{\wedge
4}p_3^+p_4^+}{K_{43}^{\wedge}K_{32}^{\wedge}K_{21}^{\wedge}K_{14}^{\wedge}p_1^+p_2^+}
\bigg[\frac{19}{9}-\frac{2}{3}\log{\delta e^\gamma{t}}
\bigg]-\frac{1}{3}{\textrm{\Large{$\times$}}}
+\frac{2}{3}\bigg\}\nn\\
\nn\\
A^{\wedge\wedge\vee\vee}_G&=&\frac{ig^4}{8\pi^2}\Tr[t^at^bt^ct^d]\bigg\{\frac{-2K_{12}^{\wedge
4}p_3^+p_4^+}{K_{43}^{\wedge}K_{32}^{\wedge}K_{21}^{\wedge}K_{14}^{\wedge}p_1^+p_2^+}\bigg[-(\log^2{\frac{s}{t}}+\pi^2)-\frac{11}{3}\log{\delta
e^\gamma{t}}+\frac{73}{9}
\bigg]-\frac{1}{3}{\textrm{\Large{$\times$}}}+\frac{2}{3}\bigg\}\nn\\
\\
\nn\\
\nn\\
A^{\wedge\vee\wedge\vee}_S&=&\frac{ig^4}{8\pi^2}\Tr[t^at^bt^ct^d]\bigg\{\frac{-2K_{13}^{\wedge
4}p_2^+p_4^+}{K_{43}^{\wedge}K_{32}^{\wedge}K_{21}^{\wedge}K_{14}^{\wedge}p_1^+p_3^+}\bigg[-\frac{s^2
t^2}{2(s+t)^4}(\log^2{\frac{s}{t}}+\pi^2)\nn\\&&+\frac{s(2t^2-5st-s^2)}{6(s+t)^3}\log{\frac{s}{t}}+\frac{1}{6}\log{\delta
e^\gamma{s}}+\frac{t s}{2(s+t)^2}+\frac{1}{18}
\bigg]-\frac{1}{6}{\textrm{\Large{$\times$}}}+\frac{1}{3}\bigg\}\nn\\
\nn\\
A^{\wedge\vee\wedge\vee}_F&=&\frac{-ig^4}{4\pi^2}\Tr[t^at^bt^ct^d]\bigg\{\frac{-2K_{13}^{\wedge
4}p_2^+p_4^+}{K_{43}^{\wedge}K_{32}^{\wedge}K_{21}^{\wedge}K_{14}^{\wedge}p_1^+p_3^+}\bigg[\frac{st(t^2+s^2)}{2(s+t)^4}(\log^2{\frac{s}{t}}+\pi^2)\nn\\&&+\frac{s(5t^2+st+2s^2)}{3(s+t)^3}\log{\frac{s}{t}}-\frac{2}{3}\log{\delta
e^\gamma{s}}+\frac{t s}{(s+t)^2}+\frac{19}{9}
\bigg]
-\frac{1}{3}{\textrm{\Large{$\times$}}}+\frac{2}{3}\bigg\}\nn\\
\nn\\
A^{\wedge\vee\wedge\vee}_G&=&\frac{ig^4}{8\pi^2}\Tr[t^at^bt^ct^d]\bigg\{\frac{-2K_{13}^{\wedge
4}p_2^+p_4^+}{K_{43}^{\wedge}K_{32}^{\wedge}K_{21}^{\wedge}K_{14}^{\wedge}p_1^+p_3^+}\bigg[-\frac{(s^2+st+t^2)^2}{(t+s)^4}(\log^2{\frac{s}{t}}+\pi^2)\nn\\&&+\frac{s}{3}\frac{(14t^2+19st+11s^2)}{(s+t)^3}\log{\frac{s}{t}}-\frac{11}{3}\log{\delta
e^\gamma{s}}+\frac{t s}{(s+t)^2}+\frac{73}{9}
\bigg]-\frac{1}{3}{\textrm{\Large{$\times$}}}+\frac{2}{3}\bigg\}\label{hc_amp}\eea

The symbol ${\textrm{\Large{$\times$}}}$ above is the relevant
four point vertex:
$-2(p_1^+p_3^++p_2^+p_4^+)/[(p_1^++p_4^+)(p_2^++p_3^+)]$ or
$2(p_2^+p_3^++p_1^+p_4^+)[(p_1^++p_2^+)(p_3^++p_4^+)]+2(p_1^+p_2^++p_3^+p_4^+)/[(p_1^++p_4^+)(p_2^++p_3^+)]$.
The four point vertex is not a valid counter term, because it is
not a polynomial of momenta. But we can add a term proportional to
$p^2$ to the self-energy term Eq.\ref{sfmass_massless}, this only
changes the field strength renormalisation by a constant, hence is
perfectly allowed. With this term, the coefficient of s and t
channel exchange diagram is shifted. So if we pick the numerical
factor in front of $p^2$ to be
$-\frac{1}{6},\;-\frac{1}{3},\;-\frac{1}{3}$ for scalar, fermion
and gluon respectively, then they will match the coefficient of
the lone four point vertex, completing it to a full tree. This
brings about a change in the numerical
factor:$\frac{1}{18}\,\rightarrow\,-\frac{5}{18}$,
$\frac{19}{9}\,\rightarrow\,\frac{13}{9}$,
$\frac{73}{9}\,\rightarrow\,\frac{67}{9}$ \cite{II}. In fact, the
numerical factor is quite unimportant, as it can be altered by a
redefinition of coupling constant.

It is tempting to go to $\mathcal{N}=4$ SYM by putting $N_s=6$,
$N_f=2$(Dirac fermions) and $N_g=1$. The result is amusing(up to
infrared terms):\bea
A^{\wedge\wedge\wedge\wedge}_{SYM}&=&A^{\wedge\wedge\wedge\vee}_{SYM}=0\nn\\
A^{\wedge\wedge\vee\vee}_{SYM}&=&\frac{ig^4}{8\pi^2}\Tr[t^at^bt^ct^d]\frac{-2K_{12}^{\wedge
4}p_3^+p_4^+}{K_{43}^{\wedge}K_{32}^{\wedge}K_{21}^{\wedge}K_{14}^{\wedge}p_1^+p_2^+}\bigg[-(\log^2{\frac{s}{t}}+\pi^2)\bigg]\nn\\
A^{\wedge\vee\wedge\vee}_{SYM}&=&\frac{ig^4}{8\pi^2}\Tr[t^at^bt^ct^d]\frac{-2K_{13}^{\wedge
4}p_2^+p_4^+}{K_{43}^{\wedge}K_{32}^{\wedge}K_{21}^{\wedge}K_{14}^{\wedge}p_1^+p_3^+}\bigg[-(\log^2{\frac{s}{t}}+\pi^2)\bigg]\eea
Quite remarkably, in this case, all counter terms cancel.

\subsection{A Word on Infrared Terms}
In the above list of results, we have omitted the infrared
sensitive terms. First, as we can see that the self-energy diagram
always contains a term $\log{\delta e^\gamma p^2x(1-x)}$. This
gives a multi-particle branch cut on the positive real axis, and
stops us from doing wave function renormalisation. This can be
cured by summing over collinear emissions or absorptions. The
analysis of \cite{II} showed that doing so is equivalent to
replacing $\log{\delta e^\gamma p^2x(1-x)}$ with $\log{\delta
e^\gamma \Delta^2x(1-x)}$, $\Delta$ being the jet resolution. For
fermion loop or scalar loop, this substitution alone is enough to
regulate the infrared divergence (the triangle or box diagrams
involving fermions or scalars are devoid of further infrared
divergences). While for gluons, we refer to \cite{II} for a
complete treatment of infrared terms. Here we only list the
infrared terms from triangle and box diagrams from the gluon loop.
The $k$ below is dual momentum, with $p_1^+=\Po-\Pt$,
$p_2^+=\Pf-\Po$, $p_3^+=\Pth-\Pf$, $p_4^+=\Pt-\Pth$ and
$\Pt<\Pth<\Po<\Pf$. $p_1$ $p_2$ are incoming legs while $p_3$
$p_4$ are outgoing.
\bea&&\frac{ig^4}{8\pi^2}\Tr[t^at^bt^ct^d]\times\nn\\
\rleft\nn\\
&&\log{\frac
{({\Pf}-\Kp)(-{\Pt}+\Kp)s\delta e^{\gamma}}{({\Pt}-{\Pf})^{2}}}\left[\frac{1}{\Kp-\Pth}+\frac{1}{\Kp-\Po}\right]\nn\\
&+&\log{\frac
{({\Po}-\Kp)(\Kp-{\Pt})s\delta e^{\gamma}}{({\Po}-{\Pt})({\Pf}-{\Pt})}}\left[-\frac{2}{\Kp-\Po}\right]\nn\\
&+&\log{\frac
{({\Pth}-\Kp)(-{\Pt}+\Kp)s\delta e^{\gamma}}{({\Pth}-{\Pt})({\Pf}-{\Pt})}}\left[-\frac{2}{\Kp-\Pth}\right]\nn\\
&+&\log{\frac
{(-{\Pt}+\Kp)^{2}t\delta e^{\gamma}}{({\Po}-{\Pt})({\Pth}-{\Pt})}}\left[\frac{2}{\Kp-\Pt}\right]\nn\\
\nn\\
\rmid\nn\\
&&\log{\frac{({\Pf}-\Kp)(-{\Pt}+\Kp)s\delta e^{\gamma}}{({\Pt}-{\Pf})^{2}}}\left[-\frac{1}{\Kp-\Pth}+\frac{1}{\Kp-\Po}\right]\nn\\
&+&\log{\frac{({\Pth}-\Kp)(\Kp-{\Po})t\delta e^{\gamma}}{(-{\Po}+{\Pth})^{2}}}\left[-\frac{1}{\Kp-\Pt}+\frac{1}{\Kp-\Pf}\right]\nn\\
&+&\log{\frac
{({\Pt}-\Kp)(\Kp-{\Po})s\delta e^{\gamma}}{(-{\Po}+{\Pt})({\Pt}-{\Pf})}}\left[-\frac{2}{\Kp-\Po}\right]\nn\\
&+&\log{\frac
{(\Kp-\Pth)({\Pf}-\Kp)s\delta e^{\gamma}}{({\Pf}-{\Pt})({\Pf}-{\Pth})}}\left[\frac{2}{\Kp-\Pth}\right]\nn\\
&+&\log{\frac
{(-{\Pf}+\Kp)({\Pth}-\Kp)t\delta e^{\gamma}}{({\Pth}-{\Pf})(-{\Po}+{\Pth})}}\left[-\frac{2}{\Kp-\Pf}\right]\nn\\
&+&\log{\frac
{({\Pt}-\Kp)(\Kp-{\Po})t\delta e^{\gamma}}{(-{\Po}+{\Pt})(-{\Po}+{\Pth})}}\left[\frac{2}{\Kp-\Pt}\right]\nn\\
\nn\\
\rright\nn\\
&&\log{\frac{({\Pf}-\Kp)(-{\Pt}+\Kp)s\delta e^{\gamma}}{({\Pt}-{\Pf})^{2}}}\left[-\frac{1}{\Kp-\Pth}-\frac{1}{\Kp-\Po}\right]\nn\\
&+&\log{\frac
{(\Kp-{\Po})(-{\Pf}+\Kp)s\delta e^{\gamma}}{({\Pf}-{\Po})({\Pt}-{\Pf})}}\left[\frac{2}{\Kp-\Po}\right]\nn\\
&+&\log{\frac
{({\Kp}-\Pth)({\Pf}-\Kp)s\delta e^{\gamma}}{({\Pf}-{\Pt})({\Pf}-{\Pth})}}\left[\frac{2}{\Kp-\Pth}\right]\nn\\
&+&\log{\frac{(-{\Pf}+\Kp)^{2}t\delta
e^{\gamma}}{({\Pf}-{\Po})({\Pf}-{\Pth})}}\left[-\frac{2}{\Kp-\Pf}\right]\nn\eea
The above is multiplied by the corresponding tree amplitude
(either $-2K_{12}^{\wedge
4}p_3^+p_4^+/(K_{43}^{\wedge}K_{32}^{\wedge}K_{21}^{\wedge}K_{14}^{\wedge}p_1^+p_2^+)$
or $-2K_{12}^{\wedge
4}p_2^+p_4^+/(K_{43}^{\wedge}K_{32}^{\wedge}K_{21}^{\wedge}K_{14}^{\wedge}p_1^+p_3^+)$
for $\wedge\wedge\vee\vee$ and $\wedge\vee\wedge\vee$).

The infrared terms above will be combined with soft Bremsstrahlung
and collinear emission (absorptions) along with the self-energy
insertions on the external legs to give a finite result. Note also
that the $\log{\delta e^{\gamma}}$ part of the above infrared
terms actually cancel the divergent x integral in the $\log{\delta
e^{\gamma}}$ part of the the gluon self-energy diagrams on the
external legs. Thus the coefficients of $\log{\delta e^\gamma}$
become $1/6$, $-2/3$ and $-11/3$(times tree) for scalar fermion
and scalar respectively. Here we quote the scattering probability
with general massless gauge and matter fields, with initial and
final states treated as in \cite{II}.

\bea
P^{\wedge\wedge\vee\vee}&=&|A^{\wedge\wedge\vee\vee}_{tree}|^2\bigg[1+\frac{g^2C(G)}{8\pi^2}\bigg[-2\log^2{\frac{\Delta^2}{s}}-2\log^2{\frac{\Delta^2}{|t|}}+\frac{2\pi^2}{3}+[\frac{67}{9}-\frac{5}{18}N_s-\frac{26}{9}N_f]\nn\\
&+&\log{\frac{\delta
e^{\gamma}\Delta^4}{|t|}}[-\frac{11}{3}+\frac{1}{6}N_s+\frac{4}{3}N_f]+\log^2{\frac{s}{|t|}}
\bigg]\bigg]\nn\\
P^{\wedge\vee\wedge\vee}&=&|A^{\wedge\vee\wedge\vee}_{tree}|^2\bigg[1+\frac{g^2C(G)}{8\pi^2}\bigg[-2\log^2{\frac{\Delta^2}{s}}-2\log^2{\frac{\Delta^2}{|t|}}+\frac{2\pi^2}{3}+[\frac{67}{9}-\frac{5}{18}N_s-\frac{26}{9}N_f]\nn\\
&+&\log{\frac{\delta e^{\gamma}\Delta^4}{s}}[-\frac{11}{3}+\frac{1}{6}N_s+\frac{4}{3}N_f]+\frac{1}{2(s+t)^4}\log^2{\frac{s}{|t|}}[s^2t^2N_s+2st(s^2+t^2)N_f+2(s^2+st+t^2)^2]\nn\\
&+&\frac{1}{6(s+t)^3}\log{\frac{s}{|t|}}[-s(2t^2-5st-s^2)N_s+4s(5t^2+st+2s^2)N_f-2s(14t^2+19st+11s^2)]\nn\\
&+&\frac{st}{2(s+t)^2}[-N_s+4N_f-2]\bigg]\bigg] \eea where
$\Delta$ is the jet resolution.

\section {Massive Matter Fields}
The main difficulty for massive field calculation is the Feynman
parameter integrals. Indeed, we can only reduce all Feynman
parameter integrals into a set of three definitive integrals. They
are: \bea
&&I(s):=\int_0^1{dx\frac{s}{sx(1-x)+M}}\nn\\
&&J(s):=\int_0^1{dx\frac{1}{x}\log{\frac{sx(1-x)+M}{M}}}\nn\\
&&K(s,t):=\int_0^1{dx\frac{1}{stx(1-x)+M(s+t)}\log{\frac{(sx(1-x)+M)(tx(1-x)+M)}{M^2}}}
\label{IJK}\eea Where $M$ is in fact $-m^2+i\epsilon$. As
$M\,\to\,0$: \bea I(s)&\sim& 2\log{\frac{s}{M}}\nn\\
J(s)&\sim& \frac{1}{2}\log^2{\frac{s}{M}}\nn\\
K(s,t)&\sim& -\pi^2+2\log{\frac{s}{M}}\log{\frac{t}{M}}\nn\\
K(s,t)-2J(s)-2J(t)&\sim& -\pi^2-\log^2{\frac{s}{t}}\eea

For example, a simple integral: \bea &&\int_0^1{
\delta(\T+\Uu+\eS+\V-1)d\T d\Uu d\eS d\V
\frac{\T\Uu(1-\T-\Uu)}{(\T\eS t+\Uu\V s +M)^2}}\nn\\
\textrm{can be reduced
to:}&&-\frac{M}{s^2(s+t)}J(s)-\frac{M}{t^2(s+t)}J(t)-\frac{M}{2st(s+t)}K(s,t)+\frac{1}{2st}
\eea

The Feynman rules for scalar fields remain good, but the
decomposition Eq.\ref{fermi_prop} does not, since it introduces
$1/q^+$ factors into the Feynman rules. This will complicate the
already complicated Feynman parameter integral. Also $q^+$ and
$q^-$ will be treated the same in contrast to the massless case,
where $q^-$ is integrated out and $q^+$ is given by
$\sum{T_ik_i^+}/\sum{T_i}$.

In order to organize the gamma matrix algebra in the fermion part
of the calculation, we make use of the factorisability of gluon
polarisation vectors Eq.\ref{polar_fact} to reduce products of
gamma matrices to products of
$K^\mu_{ij}=p_i^+p_j^\mu-p_j^+p_i^\mu$, which had been proved to
be quite handy. For example, if we are to calculate the diagram
Fig.\ref{sf_mass}. We would write down: \bea (ig)^2 \Tr(t^at^b)
\Tr\left([(q-k_0)\cdot\gamma+m]\gamma_{\mu}[(q-k_1)\cdot\gamma+m]\gamma_{\nu}\right)(-\epsilon_{\vee}^{\mu})(-\epsilon_{\wedge}^{\nu})\frac{i}{(q-k_0)^2-m^2}\frac{i}{(q-k_1)^2-m^2}\eea
The numerator can be written: \bea
&&\Tr\bigg[\left|\begin{array}{cc}
m&(q-k_0)\cdot\sigma\\(q-k_0)\cdot\bar{\sigma}&m\end{array}\right|\left|\begin{array}{cc}
0&-\sqrt{2}|\eta\rangle[k_1-k_0|\\-\sqrt{2}|k_0-k_1]\langle\eta|&0\end{array}\right|\times\nn\\
&&\left|\begin{array}{cc}
m&(q-k_1)\cdot\sigma\\(q-k_1)\cdot\bar{\sigma}&m\end{array}\right|\left|\begin{array}{cc}
0&-\sqrt{2}|k_0-k_1\rangle[\eta|\\-\sqrt{2}|\eta]\langle
k_1-k_0|&0\end{array}\right|\bigg]\nn\\
\nn\\
&&=[\eta|q-k_0|\eta\,\rangle[k_1-k_0|q-k_1|k_0-k_1\rangle+\langle
k_0-k_1|q-k_0|k_1-k_0]\,\langle\eta|q-k_1|\eta]\nn\\
&&+\langle
k_0-k_1|m|\eta\rangle\,[k_1-k_0|m|\eta]+[\eta|m|k_1-k_0]\langle
\eta|m|k_0-k_1\rangle \eea For the notation of spinor, see the
appendix. The standard procedure of momentum integration tell us
to shift $q\,\rightarrow\,q+xk_0+(1-x)k_1$, and to replace $q^\mu
q^\nu$ by ${q^2g^{\mu\nu}}/{4}$, etc. \footnote{We have omitted
the complication due to the $\delta$ regulator. For one thing, the
momentum integration is no longer homogeneous, hence the
replacement of $q^\mu q^\nu$ by ${q^2g^{\mu\nu}}/{4}$ is
problematic. Some extra work is required to sort this out, which
we will not bore the reader with.}: \bea
&&[\eta|q-k_0|\eta\,\rangle[k_1-k_0|q-k_1|k_0-k_1\rangle\nn\\
&&=[\eta|(1-x)(k_1-k_0)|\eta\,\rangle[k_1-k_0|x(k_0-k_1)|k_0-k_1\rangle+[\eta|\bar{\sigma}^\mu_{\dot{a}a}|\eta\,\rangle[k_1-k_0|\bar{\sigma}^\nu_{\dot{b}b}|k_0-k_1\rangle\frac{g_{\mu\nu}q^2}{4}\nn\\
&&=(1-x)\sqrt{2}(k_1-k_0)^+\,\sqrt{2}\frac{x(k_0-k_1)^2}{2(k_0-k_1)^+}+[\eta|k_1-k_0]\langle
k_0-k_1|\eta\rangle\,\frac{q^2}{2}\nn\\
&&=-x(1-x)(k_0-k_1)^2+\frac{q^2}{2}\eea here we have used an
identity:
\bea\bar{\sigma}^{\mu}_{\dot{a}a}\bar{\sigma}^{\nu}_{\dot{b}b}g_{\mu\nu}=2\epsilon_{\dot{a}\dot{b}}\epsilon_{ab};\
\ \
\sigma_{\mu}^{a\dot{a}}\sigma_{\nu}^{b\dot{b}}g^{\mu\nu}=2\epsilon^{ab}\epsilon^{\dot{a}\dot{b}}\eea

For a more complicated example, a string of spinor products
becomes,
\bea&&[p_1|q-k_0|p_4\rangle[\eta|q-k_3|p_3\rangle\nn\\
&=&[p_1|q+\T(k_1-k_0)+\V(k_2-k_0)+\eS(k_3-k_0)|p_4\rangle[\eta|q+\Uu(k_0-k_3)+\T(k_1-k_3)+\V(k_2-k_3)|p_3\rangle\nn\\
&=&[p_1|\eta]\langle p_3|p_4\rangle\frac{q^2}{2}+[p_1|\V p_2|p_4\rangle[\eta|\Uu p_4-\T p_2|p_3\rangle\nn\\
&=&(-1)\frac{1}{p_3^+p_4^+}K^{\wedge}_{43}\frac{q^2}{2}+\frac{\sqrt{2}\V}{p_1^+p_2^+p_4^+}K^{\vee}_{12}K^{\wedge}_{42}(\frac{\sqrt{2}\Uu}{p_3^+}K^{\wedge}_{34}-\frac{\sqrt{2}\T}{p_3^+}K^{\wedge}_{32})\nn\eea

Basically, all of the spinor products can be reduced to one of the
$K_{ij}$'s, and the reader can find in the appendix some practical
details as to how to organize the products of $K_{ij}$'s. After
the momentum integral is done, we can perform the Feynman
parameter integrals using Eq.\ref{IJK}.

\section {Results}
\subsection{Self-energy Diagrams}
For Fig.\ref{sf_mass}, the results are: \bea
&&\Pi^{\wedge\vee}_S=-\frac{ig^2}{16\pi^2}\Tr[t^at^b]\bigg[\frac{1}{6}p^2\log{\delta
e^\gamma
m^2}+\frac{1}{12}\frac{(p^2-4m^2)^2}{p^2}I(p^2)-\frac{5}{18}p^2+\frac{4}{3}m^2\bigg]\nn\\
&&\Pi^{\wedge\vee}_F=\frac{ig^2}{8\pi^2}\Tr[t^at^b]\bigg[-\frac{2}{3}p^2\log{\delta
e^\gamma
m^2}-\frac{1}{3}\frac{(p^2-4m^2)(p^2+2m^2)}{p^2}I(p^2)+\frac{13}{9}p^2+\frac{8}{3}m^2\bigg]\eea
The gluon mass counter term in the expressions above has been
removed already.

\subsection{Triangle Diagrams}
The results for triangle diagram Fig.\ref{trig} are: \bea
\Gamma^{\wedge\wedge\vee}_S&=&\frac{ig^3}{8\pi^2}\Tr[t^at^bt^c]\times\frac{-2p_3^+}{p_1^+p_2^+}K^\wedge_{21}\bigg[\frac{1}{6}\log{m^2\delta
e^{\gamma}}+(p_o^2-4m^2)(\frac{(p_o^2-4m^2)}{12p_o^4}+\alpha\frac{p_1^+p_2^+}{p_3^{+2}}\frac{m^2}{p_o^4})I(p_o^2)\nn\\&&
-\alpha\frac{p_1^+p_2^+}{p_3^{+2}}\frac{m^2}{p_o^2}J(p_o^2)-\frac{1}{9}+\frac{4m^2}{3p_o^2}-\alpha\frac{1}{6}\frac{p_1^+p_2^+}{p_3^{+2}}(1+\frac{24m^2}{p_o^2})\bigg]\nn\\
\Gamma^{\wedge\wedge\vee}_F&=&-\frac{ig^3}{4\pi^2}\Tr[t^at^bt^c]\times\frac{-2p_3^+}{p_1^+p_2^+}K^\wedge_{21}\bigg[-\frac{2}{3}\log{m^2\delta
e^{\gamma}}+(p_o^2-4m^2)(-\frac{(p_o^2+2m^2)}{3p_o^4}+2\alpha\frac{p_1^+p_2^+}{p_3^{+2}}\frac{m^2}{p_o^4})I(p_o^2)\nn\\
&&-2\alpha\frac{p_1^+p_2^+}{p_3^{+2}}\frac{m^2}{p_o^2}J(p_o^2)+\frac{16}{9}+\frac{8m^2}{3p_o^2}-\alpha\frac{1}{3}\frac{p_1^+p_2^+}{p_3^{+2}}(1+\frac{24m^2}{p_o^2})\bigg]\eea
where $\alpha=1$ if leg 3 is off shell and 0 otherwise, and $p_o$
is the off shell momentum. Their anomalous terms are identical
with the massless result, which agrees with the fact that
anomalous terms are UV effects.

The result for the MHV triangle is: \bea
\Gamma^{\wedge\wedge\wedge}_S&=&\frac{ig^3}{8\pi^2}\Tr[t^at^bt^c]\frac{(K^{\wedge}_{21})^3}{p_1^+p_2^+p_3^+}\bigg[-\frac{4m^2(p_o^2-4m^2)}{p_o^6}I(p_o^2)+\frac{4m^2}{p_o^4}J(p_o^2)+\frac{2(p_o^2+24m^2)}{3p_o^4}\bigg]\nn\\
\Gamma^{\wedge\wedge\wedge}_F&=&-\frac{ig^3}{4\pi^2}\Tr[t^at^bt^c]\frac{(K^{\wedge}_{21})^3}{p_1^+p_2^+p_3^+}\bigg[-\frac{8m^2(p_o^2-4m^2)}{p_o^6}I(p_o^2)+\frac{8m^2}{p_o^4}J(p_o^2)+\frac{4(p_o^2+24m^2)}{3p_o^4}\bigg]\eea

\subsection{Scattering Amplitudes}

\bea&&A^{\wedge\wedge\wedge\wedge}_S=\frac{ig^4}{\pi^2}\Tr[t^at^bt^ct^d]\frac{K_{43}^{\wedge}K_{32}^{\wedge}K_{21}^{\wedge}K_{14}^{\wedge}}{p_1^+p_2^+p_3^+p_4^+st}\bigg[\frac{1}{6}-\frac{m^4}{st}K(s,t)\bigg]\nn\\
&&A^{\wedge\wedge\wedge\wedge}_F\frac{-2ig^4}{\pi^2}\Tr[t^at^bt^ct^d]\frac{K_{43}^{\wedge}K_{32}^{\wedge}K_{21}^{\wedge}K_{14}^{\wedge}}{p_1^+p_2^+p_3^+p_4^+st}\bigg[\frac{1}{3}-\frac{2m^4}{st}K(s,t)\bigg]\eea

\bea
A^{\wedge\wedge\wedge\vee}_S&=&\frac{ig^4}{8\pi^2}\Tr[t^at^bt^ct^d]\frac{K_{13}^{\wedge
2}p_2^+p_4^+}{K_{43}^{\wedge}K_{32}^{\vee}K_{21}^{\vee}K_{14}^{\wedge}}\times\nn\\
&&\bigg[-\frac{(2t+s)tm^2}{2(s+t)s}(1-\frac{4m^2}{s})I(s)-\frac{(2s+t)sm^2}{2(s+t)t}(1-\frac{4m^2}{t})I(t)\nn\\
&&+\frac{(2s+t)t^2m^2}{(s+t)^2s}J(s)+\frac{(2t+s)s^2m^2}{(s+t)^2t}J(t)+\frac{stm^2}{2(s+t)^2}(1-2\frac{m^2(s+t)}{st})K(s,t)\nn\\
&&+\frac{2(2s^2-st+2t^2)m^2}{st}+\frac{(s+t)}{6}\bigg]\nn\\
\nn\\
A^{\wedge\wedge\wedge\vee}_F&=&\frac{-ig^4}{4\pi^2}\Tr[t^at^bt^ct^d]\frac{K_{13}^{\wedge
2}p_2^+p_4^+}{K_{43}^{\wedge}K_{32}^{\vee}K_{21}^{\vee}K_{14}^{\wedge}}\times\nn\\
&&\bigg[-\frac{(2t+s)tm^2}{(s+t)s}(1-\frac{4m^2}{s})I(s)-\frac{(2s+t)sm^2}{(s+t)t}(1-\frac{4m^2}{t})I(t)\nn\\
&&+\frac{2(2s+t)t^2m^2}{(s+t)^2s}J(s)+\frac{2(2t+s)s^2m^2}{(s+t)^2t}J(t)+\frac{stm^2}{(s+t)^2}(1-2\frac{m^2(s+t)}{st})K(s,t)\nn\\
&&+\frac{4(2s^2-st+2t^2)m^2}{st}+\frac{(s+t)}{3}\bigg]\eea

In contrast to the massless case, the external leg factors are
included in the results below for the helicity conserving
amplitude.
\bea
A^{\wedge\wedge\vee\vee}_S&=&\frac{ig^4}{8\pi^2}\Tr[t^at^bt^ct^d]\frac{-2K_{12}^{\wedge
4}p_3^+p_4^+}{K_{43}^{\wedge}K_{32}^{\wedge}K_{21}^{\wedge}K_{14}^{\wedge}p_1^+p_2^+}\times\bigg\{\nn\\
&&\bigg[(\frac{m^2}{2s}-\frac{2m^2}{3t}-\frac{2m^4}{st}+\frac{4m^4}{3t^2}+\frac{1}{12})I(t)+\frac{m^4}{s^2}K(s,t)-\frac{2m^2}{s}+\frac{4m^2}{3t}+\frac{1}{18}\nn\\
&&-\frac{1}{6}\log{\delta e^\gamma
m^2}-\frac{1}{3}\bigg]-\frac{1}{6}{\textrm{\Large{$\times$}}}+\frac{1}{3}\bigg\}
\nn\\
A^{\wedge\wedge\vee\vee}_F&=&\frac{-ig^4}{4\pi^2}\Tr[t^at^bt^ct^d]\frac{-2K_{12}^{\wedge
4}p_3^+p_4^+}{K_{43}^{\wedge}K_{32}^{\wedge}K_{21}^{\wedge}K_{14}^{\wedge}p_1^+p_2^+}\nn\\
&&\bigg[(\frac{m^2}{s}+\frac{2m^2}{3t}-\frac{4m^4}{st}+\frac{8m^4}{3t^2}-\frac{1}{3})I(t)+(-\frac{m^2}{s}+\frac{2m^4}{s^2})K(s,t)-\frac{4m^2}{s}+\frac{8m^2}{3t}+\frac{19}{9}\nn\\
&&+\frac{2}{3}\log{\delta e^\gamma
m^2}-\frac{2}{3}\bigg]-\frac{1}{3}{\textrm{\Large{$\times$}}}+\frac{2}{3}\bigg\}\eea

\bea
A^{\wedge\vee\wedge\vee}_S&=&\frac{ig^4}{8\pi^2}\Tr[t^at^bt^ct^d]\frac{-2K_{13}^{\wedge
4}p_2^+p_4^+}{K_{43}^{\wedge}K_{32}^{\wedge}K_{21}^{\wedge}K_{14}^{\wedge}p_1^+p_3^+}\nn\\
&&\bigg[(-\frac{tm^2(17st+s^2+4t^2)}{6(s+t)^3s}+\frac{2tm^4(2t+5s)}{3(s+t)^2s^2}+\frac{t(5st-2s^2+t^2)}{12(s+t)^3})I(s)\nn\\
&&+(-\frac{sm^2(17st+t^2+4s^2)}{6(s+t)^3t}+\frac{2sm^4(5t+2s)}{3(s+t)^2t^2}+\frac{s(5st-2t^2+s^2)}{12(s+t)^3})I(t)\nn\\
&&+(\frac{2stm^2}{(s+t)^3}-\frac{s^2t^2}{(s+t)^4})(J(s)+J(t))+(-\frac{2stm^2}{(s+t)^3}+\frac{m^4}{(s+t)^2}+\frac{s^2t^2}{2(s+t)^4})K(s,t)\nn\\
&&+\frac{2m^2(2t^2+2s^2+3st)}{3(s+t)st}+\frac{s^2+t^2+11st}{18(s+t)^2}\nn\\
&&-\frac{1}{6}\log{\delta e^\gamma
m^2}-\frac{1}{3}\bigg]-\frac{1}{6}{\textrm{\Large{$\times$}}}+\frac{1}{3}\bigg\}\nn\\
A^{\wedge\vee\wedge\vee}_F&=&\frac{-ig^4}{4\pi^2}\Tr[t^at^bt^ct^d]\frac{-2K_{13}^{\wedge
4}p_2^+p_4^+}{K_{43}^{\wedge}K_{32}^{\wedge}K_{21}^{\wedge}K_{14}^{\wedge}p_1^+p_3^+}\nn\\
&&\bigg[(\frac{tm^2(5s^2+2t^2-5st)}{3(s+t)^3s}+\frac{4tm^4(2t+5s)}{3(s+t)^2s^2}-\frac{t(5s^2+2t^2+st)}{6(s+t)^3})I(s)\nn\\
&&+(\frac{tm^2(5t^2+2s^2-5st)}{3(s+t)^3t}+\frac{4sm^4(5t+2s)}{3(s+t)^2t^2}-\frac{s(5t^2+2s^2+st)}{6(s+t)^3})I(t)\nn\\
&&+(\frac{4stm^2}{(s+t)^3}+\frac{st(s^2+t^2)}{(s+t)^4})(J(s)+J(t))+(\frac{m^2(s-t)^2}{(s+t)^3}+\frac{2m^4}{(s+t)^2}-\frac{st(s^2+t^2)}{2(s+t)^4})K(s,t)\nn\\
&&+\frac{4m^2(2t^2+2s^2+3st)}{3(s+t)st}+\frac{19s^2+19t^2+47st}{9(s+t)^2}\nn\\
&&+\frac{2}{3}\log{\delta e^\gamma
m^2}-\frac{2}{3}\bigg]-\frac{1}{3}{\textrm{\Large{$\times$}}}+\frac{2}{3}\bigg\}\eea
The last numerical factors $-1/3$ and $-2/3$ inside each square
bracket is due to the amputation of external legs: $[\lim_{p^2\to
0}{\Pi(p^2)/p^2}]^{1/2}$. 

\section{Concluding Remarks}
In summary, we have calculated the next to leading order gluon
scattering amplitudes with fermions and scalars in the loop, as an
extension of \cite{I,II}. They are all part of the project of
showing that summing planar diagrams on the light cone world sheet
does give correct results (at least at one loop order). The light
cone world sheet tells us to organize diagrams such that an
explicit cancellation of gauge artificial divergence can be seen.
There is no gauge artificial divergences in the fermion and scalar
part of the calculation, and the method developed in \cite{II}
works out as expected. And the same counter terms required in
\cite{II} appear here too.

In the special case of $\mathcal{N}=4$ SYM, all of these counter
terms cancel among different sectors. This is a mere additional
bonus from the field theory point of view, but quite crucial for
the project proposed in \cite{I,II}. More specifically, in
\cite{II}, the dimension of the target space was increased to
accommodate extra world sheet fields (spurions), these fields
couple to gluons in such a way that will produce the counter terms
needed (the four point contact vertex and a 2/3 in
Eq.\ref{hc_amp}). In order to formulate $\mathcal{N}=4$ SYM on the
world sheet we need to increase the space time dimension to ten,
so as to have six more world sheet fields in addition to the
original $\mathbf{q}^{i}(\sigma,\tau),\ i=1,2$
\cite{gudmundssontt}. These eight fields correspond to the eight
bosonic degrees of freedom in $\mathcal{N}=4$ SYM. The AdS/CFT
duality says this theory is dual to the string theory in 10
dimensions, meaning we no longer have room for extra dimensions to
produce counter terms.

The massive matter contribution to the gluon scattering was also
calculated for completeness. An exponential damping factor was
used instead of the dimension reduction as UV regulator, so the
box integral is slightly different from
\cite{Ussy_Davy,ph_ph_scatt}, so no machinery of hypergeometric
functions need be used.

\section{Acknowledgements}
The author would like to thank C.B.Thorn for carefully reading the
manuscripts and giving a lot of valuable advices. This work is
supported in part by the Department of Energy under Grant No.
DE-FG02-97ER-41029 and the Alumni Fellowship of University of
Florida.
\appendix
\section{Notation}
\bea&&\gamma^\mu=\left|\begin{array}{cc}0&\sigma^\mu\\\bar{\sigma}^\mu&0\end{array}\right|\ \ \sigma^\mu:=(\textrm{I},\vec{\sigma})\ \ \ \bar{\sigma}^\mu:=(\textrm{I},-\vec{\sigma})\nn\\
&&\epsilon_{ab}=\epsilon^{ab}=\epsilon^{\dot{a}\dot{b}}=\epsilon_{\dot{a}\dot{b}}=i\sigma_2\nn\\
&&p_a=\epsilon_{ab}p^b\ \ p^a=p_b\epsilon^{ba}\nn\\
&&|p\,]:=p_{\dot{a}}\ \ |p\,\rangle:=p^a\ \ [p|:=p^{\dot{a}}\ \
\langle p|:=p_a\nn\\
&&\eta^\alpha=\eta^{\dot{\alpha}}=\left|\begin{array}{c}1\\0\end{array}\right|\
\
\eta_\alpha=\eta_{\dot{\alpha}}=\left|\begin{array}{c}0\\-1\end{array}\right|\eea

We can define the light cone version spinor as: \bea
&&p^{\alpha}=\left|\begin{array}{c}-\frac{p^\wedge}{p^+}\\1\end{array}\right|\hspace{.2in}p_{\dot\alpha}=\left|\begin{array}{c}1\\\frac{p^\vee}{p^+}\end{array}\right|\label{spinor}\eea
The spinors satisfy the Dirac equation if $p$ is light like. Note
that they don't have the correct normalisation, namely $p\cdot
\sigma^{\alpha \dot{\alpha}}\ne p^{\alpha} p^{\dot{\alpha}}$, but
they have the merit that $p^{\alpha}=(-p)^{\alpha}$. The
polarisation vectors of gluon can be written as: \bea
&&\epsilon_{\wedge\dot{a}a}=\sqrt{2}\frac{|p\,]\langle
\eta\,|}{[\eta\,|p\,]}=-\sqrt{2}|p\,]\langle \eta\,|,\ \
\epsilon_{\vee\dot{a}a}=\sqrt{2}\frac{|\eta\,]\langle
p\,|}{[p\,|\eta\,]}=-\sqrt{2}|\eta\,]\langle p\,|\label{polar_fact}\\
&&K^\wedge_{ij}=p_j^+ p_i^+\,\langle p_j\,|p_i\,\rangle,\ \
K^\vee_{ij}=p_i^+ p_j^+[p_i\,|p_j\,]\nn\eea The $K_{ij}$'s
satisfy:
\bea \sum_j K^\mu_{ij}&=&0\nn\\
p^+_iK^\mu_{jk}+p^+_kK^\mu_{ij}+p^+_jK^\mu_{ki}&=&0\nn\\
K^\wedge_{li}K^\wedge_{jk}+K^\wedge_{lk}K^\wedge_{ij}
+K^\wedge_{lj}K^\wedge_{ki}&=&0\nn\\
\sum_j{K^\wedge_{ij}K^\vee_{jk}\over p^+_j}&=&p^+_ip^+_k\sum_j
{p_j^2\over2p^+_j} \eea

In the current case we are dealing with, $i,j$ run from 1 to 4,
but only two of the six $K_{ij}$'s are independent, say $K_{43}$
and $K_{32}$. And any product of $K_{ij}$'s with total helicity 4
can be reduced to either $(K^\wedge_{43})^4$ or
$(K^\wedge_{43})^3K^\wedge_{32}$. Product of helicity 2 can be
reduced to $(K^\wedge_{43})^2$ and $K^\wedge_{43}K^\wedge_{32}$.
Product of helicity 0 can be reduced to $1$ and
$K^\wedge_{43}K^\vee_{32}$. The reduction is in general a
formidable task for human, but quite a piece of cake for
computers, as all our calculations are done with computers.
\section{Feynman Rules}
\begin{figure}[!h]
\begin{center}
\psfrag{'mu','a'}{$\wedge,\;a$} \psfrag{'nu','a'}{$\vee,\;a$}
\psfrag{'q'}{$q$}\psfrag{'c','P3'}{$c,\;p_3$}\psfrag{'b','P2'}{$b,\;p_2$}
\includegraphics[width=2.5in]{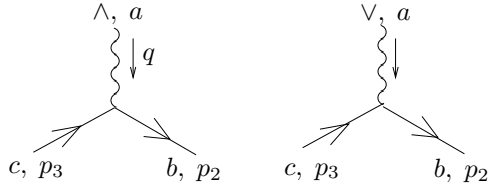}
\caption{gluon-fermion-fermion 3 point vertex}\label{gff1_eps}
\end{center}
\end{figure}
The fermion gluon vertex Fig.\ref{gff1_eps} is given by:
\bea&&V^{\wedge}_{l}=-2ig (t^a)_{bc}
\frac{p_2^{+}}{q^+p_1^+}K^{\wedge}_{p_1,q}\hspace{.5in}V^{\wedge}_{r}=-2ig
(t^a)_{bc} \frac{1}{q^+}K^{\wedge}_{p_2,q}\nn\\
&&V^{\vee}_{l}=-2ig (t^a)_{bc}
\frac{1}{q^+}K^{\vee}_{p_2,q}\hspace{.5in}V^{\vee}_{r}=-2ig
(t^a)_{bc} \frac{p_2^{+}}{q^+p_1^+}K^{\vee}_{p_1,q}\eea where the
subscript $l,\,r$ denotes whether the fermion is left or right
handed.

\begin{figure}[!h]
\begin{center}
\psfrag{'mu', 'a', 'q'}{$\mu,\;a,\;q$}\psfrag{ 'c',
'pc'}{$c,\;p_c$} \psfrag{ 'b', 'pb'}{$b,\;p_b$}
\includegraphics[height=1.2in]{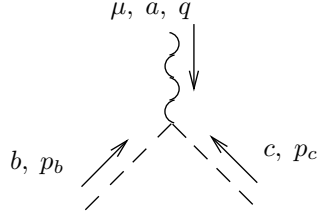}
\caption{gluon-scalar-scalar 3 point
vertex}\label{gss1_eps}\end{center}
\end{figure}
The gluon scalar vertex Fig.\ref{gss1_eps} is given by: \bea
V^{\wedge(\vee)}=-2ig (t^a)_{bc}
\frac{1}{p_b^++p_c^+}K^{{\wedge(\vee)}}_{p_c,p_b}\eea Here we have
used real scalar fields and hence it transforms in a real
representation.

\begin{figure}[!h]
\begin{center}
\psfrag{'e'}{$e$}\psfrag{'b','P2'}{$b,\;p_1$}\psfrag{'c','P3'}{$c,\;p_2$}\psfrag{'mu','a','P1'}{$a,\;p_4$}\psfrag{'nu','d','P4'}{$d,\;p_3$}
\includegraphics[width=3in]{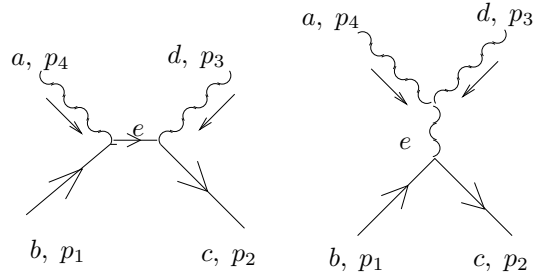}
\caption{Two diagrams contribute to the fermion-gluon 4 point
vertex. Note these are not exchange diagrams but rather four point
vertices gotten through cancelling a ${1}/{p^2}$ factor in a
fermion propagator or a gluon propagator, see Eq.\ref{fermi_prop}}
\label{ggff1_eps}
\end{center}
\end{figure}
The gluon scalar vertex Fig.\ref{ggff1_eps} is given by:
\bea&&V^{\wedge\vee}_l=-2ig^2
(t^d)_{ce}(t^a)_{eb}\frac{p_2^{+}}{p_1^++p_4^+}+2g^2 f^{dae}
(t^e)_{cb}\frac{(p_3^{+}-p_4^{+})p_2^+}{(p_3^++p_4^+)^2}\nn\\
&&V^{\wedge\vee}_r=2g^2 f^{dae}
(t^e)_{cb}\frac{(p_3^{+}-p_4^{+})p_2^+}{(p_3^++p_4^+)^2}\nn\\
&&V^{\vee\wedge}_l=2g^2 f^{dae}
(t^e)_{cb}\frac{(p_3^{+}-p_4^{+})p_2^+}{(p_3^++p_4^+)^2}\nn\\
&&V^{\vee\wedge}_r=-2ig^2
(t^d)_{ce}(t^a)_{eb}\frac{p_2^{+}}{p_1^++p_4^+}+2g^2 f^{dae}
(t^e)_{cb}\frac{(p_3^{+}-p_4^{+})p_2^+}{(p_3^++p_4^+)^2}\eea where
the first super index on $V$ refer to the polarisation of $p_3$,
the second $p_4$.

\begin{figure}[!h]
\begin{center}
\psfrag{'nu','b','p2'}{$\wedge,\;b,\;p_4$}\psfrag{'mu','a','p1'}{$\vee,\;a,\;p_3$}
\psfrag{'c','p3'}{$c,\;p_1$}\psfrag{'d','p4'}{$d,\;p_2$}
\includegraphics[width=3in]{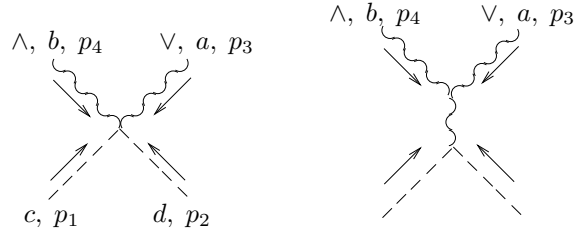}
\caption{Scalar-gluon 4 point vertex. Again, the second diagram is
not an exchange diagram but rather a four point vertex obtained
through shrinking a gluon propagator} \label{ggss_eps}
\end{center}
\end{figure}
The gluon scalar four point vertex Fig.\ref{ggss_eps} is given by:
\bea&&V_{ggss}=-ig^2\left[(t^a)_{ce}(t^b)_{ed}+(t^b)_{ce}(t^a)_{ed}\right]+g^2f^{abe}(t^e)_{cd}\frac{(p_3^{+}-p_4^{+})(p_2^+-p_1^+)}{(p_1^++p_2^+)^2}\eea

\begin{figure}[!h]
\begin{center}
\psfrag{nu,b,P2}{$\nu,\;b,\;p_2$}\psfrag{mu,a,P1}{$\mu,\;a,\;p_1$}
\psfrag{rho,c,P3}{$\rho,\;c,\;p_3$}
\includegraphics[width=1.3in]{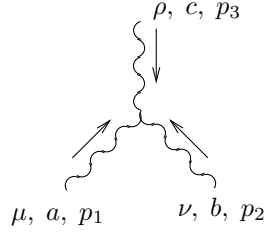}\caption{Tri-gluon vertex}
\end{center}
\end{figure}
For tri-gluon vertex Fig.\ref{ggg}  \bea V_{ggg}=
-gf^{abc}[\epsilon_1^*\cdot\epsilon_2^*(p_1-p_2)\cdot\epsilon_3^*+\epsilon_2^*\cdot\epsilon_3^*(p_2-p_3)\cdot\epsilon_1^*+\epsilon_3^*\cdot\epsilon_1^*(p_3-p_1)\cdot\epsilon_2^*]\nn\eea
Setting $\mu,\nu=\wedge$, $\rho=\vee$, the above becomes: \bea
&&gf^{abc}[(p_2-p_3)^+\frac{p_1^\wedge}{p_1^+}-(p_2-p_3)^\wedge+(p_3-p_1)^+\frac{p_2^\wedge}{p_2^+}-(p_3-p_1)^\wedge]\nn\\
&&=2gf^{abc}\frac{(p_1+p_2)^+}{p_1^+p_2^+}K_{21}^\wedge\eea
\begin{figure}[!h]
\begin{center}
\psfrag{nu,b,P2}{$\nu,\;b,\;p_2$}\psfrag{mu,a,P1}{$\mu,\;a,\;p_1$}
\psfrag{rho,c,P3}{$\rho,\;c,\;p_3$}\psfrag{sigma,d,P4}{$\sigma,\;d,\;p_4$}
\psfrag{nu1,b,P2}{$\nu,\;b,\;p_2$}\psfrag{mu1,a,P1}{$\mu,\;a,\;p_1$}
\psfrag{rho1,c,P3}{$\rho,\;c,\;p_3$}\psfrag{sigma1,d,P4}{$\sigma,\;d,\;p_4$}
\includegraphics[width=3in]{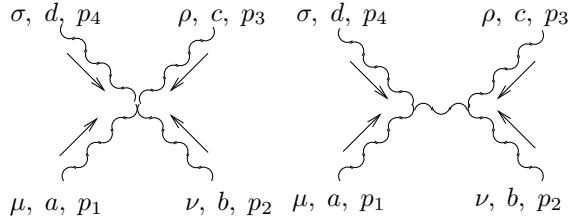}\caption{Gluon 4 point vertex}
\label{gggg}
\end{center}
\end{figure}

The gluon four point vertex receives contribution from two
sources: the left diagram in Fig.\ref{gggg} is simply the
covariant four point vertex: \bea
V_{1}=ig^2f^{abe}f^{ecd}[g^{\mu\sigma}g^{\nu\rho}-g^{\mu\rho}g^{\nu\sigma}]+ig^2f^{dae}f^{ebc}[g^{\mu\nu}g^{\rho\sigma}-g^{\mu\rho}g^{\nu\sigma}]+ig^2f^{cae}f^{ebd}[g^{\mu\nu}g^{\rho\sigma}-g^{\mu\sigma}g^{\nu\rho}]\eea
The second is obtained by shrinking a propagator. \bea
V_{2}&=&g^2f^{dae}f^{ebc}[\epsilon_1^*\cdot\epsilon_4^*(p_4-p_1){_\alpha}]\frac{ig^{\alpha+}g^{\beta+}(p_1+p_4)^2}{(p_1^++p_4^+)^2(p_1+p_4)^2}[\epsilon_2^*\cdot\epsilon_3^*(p_2-p_3){_\beta}]\nn\\
&=&ig^2f^{dae}f^{ebc}\frac{(\epsilon_1^*\cdot\epsilon_4^*)(\epsilon_2^*\cdot\epsilon_3^*)(p_4-p_1)^+(p_2-p_3){^+}}{(p_1^++p_4^+)^2}\nn\eea
There are two cases in which the gluon four point vertex is
nonzero: \bea
&&\,V_{\wedge\wedge\vee\vee}=-2ig^2f^{dae}f^{ebc}\frac{p_1^+p_3^++p_2^+p_4^+}{(p_1^++p_4^+)(p_2^++p_3^+)}+2ig^2f^{ace}f^{ebd}\frac{p_3^+p_2^++p_1^+p_4^+}{(p_1^++p_3^+)(p_2^++p_4^+)}\nn\\
&&\,V_{\wedge\vee\wedge\vee}=2ig^2f^{abe}f^{ecd}\frac{p_2^+p_3^++p_1^+p_4^+}{(p_1^++p_2^+)(p_3^++p_4^+)}+2ig^2f^{dae}f^{ebc}\frac{p_1^+p_2^++p_3^+p_4^+}{(p_1^++p_4^+)(p_2^++p_3^+)}\eea
When using these vertices, we need to watch the indices of
structure constants closely: not all terms are going to make
contributions to $\Tr[t^at^bt^ct^d]$.

\end{document}